# Why do surface tensions of most of organic liquids demonstrate close values?

Edward Bormashenko

*Ariel University Center of Samaria, Applied Physics Department, P.O.B. 3, Ariel, 40700, Israel*

e-mail: edward@ariel.ac.il

**Abstract**

Values of surface tension of most of organic liquids are of the same order of magnitude. The natural explanation for this lies in the fact that surface tension is governed by London dispersion forces, which are independent of the permanent dipole moment of molecules. The surface tension of organic liquids (with the exception of polymers and polymer solutions) depends on the ionization potential and the diameter of the molecule only. These parameters vary slightly for organic liquids.

**I. INTRODUCTION**

Surface tension is one of the most fundamental properties of liquid and gaseous phases.[1-4] Surface tension governs many phenomena in climate formation, plant biology, medicine, etc. The origin of surface tension is related to the unusual energetic state of the surface molecule, which misses about half its interactions (see Fig. 1). The similar values of surface tensions of liquids that are very different in their physical and chemical nature summarized in Table I catch the eye.[5] Indeed, the values of surface tension of most of organic liquids are located in the narrow range of 20–65 mJ/m$^2$. This is in striking contrast to other mechanical properties of liquids, such as viscosity. For example, the viscosity of ethyl alcohol at ambient conditions equals 1.2 10$^{-3}$ kg/ms, whereas the viscosity of glycerol is 1.5 kg/ms, at the same time the surface tensions of alcohol and glycerol are of the same order of magnitude.[5] The more striking example is honey, the viscosity of which may be very high, however its surface tension is 50–60 mJ/m$^2$.[6] The reasonable question is: Why is the range of values of surface tension so narrow? My paper will try to answer this question in the spirit of the famous manuscript of V. Weisskopf, who explained why materials are as dense and heavy as they are.[7] The explanation will lead to the expression relating surface tension to fundamental physical constants.

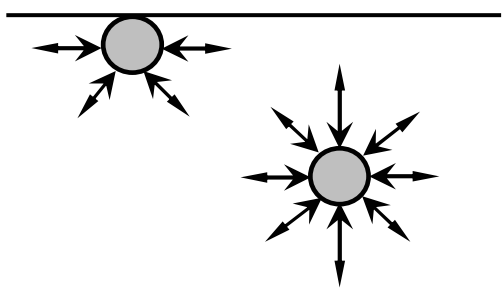

Fig. 1. A molecule at the surface misses about half its interactions.

Table I. Surface Tension, Enthalpy of Vaporization and Dipole Moment of Some of Organic Molecules.

| Liquid | Surface Tension, $\gamma$, mJ/m$^2$ | Enthalpy of Vaporization, $\Delta H$, kJ/mol | Dipole Moment, $p$, D[*] |
|---|---|---|---|
| Glycerol, $C_3H_8O_3$ | 64.7 | 91.7 | 2.56 |
| Formamide, $CH_3ON$ | 55.5 | 60.0 | 3.7 |
| $CCl_4$ | 25.7 | 32.54 | 0 |
| Chloroform, $CHCl_3$ | 26.2 | 31.4 | 1.04 |
| Dichloromethane, $CH_2Cl_2$ | 31 | 28.6 | 1.60 |
| Toluene, $C_7H_8$ | 28.5 | 38.06 | 0.36 |
| Ethyl alcohol, $C_2H_6O$ | 22 | 38.56 | 1.7 |
| Acetone, $C_3H_6O$ | 24 | 31.3 | 2.9 |

[*]The unit of a dipole moment is Debye: $1\,\text{D} = 3.3 \cdot 10^{-30}\,\text{C} \cdot \text{m}$.

## II. SURFACE TENSION AND INTERMOLECULAR FORCES

The energy states of molecules in the bulk and at the surface of liquid are not the same due to the difference in the nearest surrounding of a molecule. Each molecule in the bulk is surrounded by others on every side, whereas, for the molecule located at the liquid/vapor interface, there are very few molecules outside, as shown in Fig. 1. An important misinterpretation should be avoided: the resulting force acting on the molecule in the bulk and at the interface equals zero (both "bulk" and "interface" molecules are in mechanical equilibrium). This abundant misinterpretation was revealed and analyzed in Ref. 8. However, an increase in the liquid/vapor surface causes a rise in the quantity of "interface" molecules and consequent growth in the surface energy. Liquids tend to diminish the number of interface molecules to decrease surface energy. We can measure the surface tension by performing work when bringing molecules from the interior to the surface. Surface tension could be defined in two parallel ways: as work necessary to increase the surface area, or as a force along a line of unit length, where the force is parallel to the surface but perpendicular to the line. Let the potential describing the pair intermolecular interaction in the liquid be $U(r)$. The surface tension $\gamma$ could be estimated as:

$$\gamma = f_m \frac{1}{d_m} \cong \frac{N}{2} \frac{|U(d_m)|}{d_m} \frac{1}{d_m} = \frac{N}{2} \frac{|U(d_m)|}{d_m^2}, \qquad (1)$$

where $f_m$ is the force necessary to bring a molecule to the surface, which could be roughly estimated as $f_m \cong (N/2)|U(d_m)|/d_m$, where $d_m$ is the diameter of the molecule, $N$ – is the number of nearest neighbor molecules (the multiplier 1/2 is due to the absence of molecules "outside", i.e., in the vapor phase), and $1/d_m$ is the number of molecules per unit length of the liquid surface.

The main question is: what is the physical nature of the intermolecular potential $U(r)$? In general, there are three main kinds of intermolecular interactions:

1) The attractive interaction between identical dipolar molecules given by the Keesom formula:

$$U_{attr}(r) = -\frac{p^4}{3(4\pi\varepsilon_0)^2 kT}\frac{1}{r^6}, \qquad (2)$$

where $p$ is the dipole moment of the molecule, $k$ is the Boltzmann constant, $T$ is the temperature and $r$ is the distance between molecules.[3]

2) The Debye attractive interaction between dipolar molecules and induced dipolar molecules is:

$$U_{attr}(r) = -\frac{2p^2\alpha}{(4\pi\varepsilon_0)^2}\frac{1}{r^6}, \qquad (3)$$

where $\alpha$ is the polarizability of the molecule.[3]

3) The London dispersion interactions which are of a pure quantum mechanical nature. The London dispersion force is an attractive force that results when the electrons in two adjacent atoms occupy positions that make the atoms form temporary dipoles; its potential is given by:

$$U_{attr}(r) = -\frac{3\alpha^2 I}{4(4\pi\varepsilon_0)^2}\frac{1}{r^6}, \qquad (4)$$

where $I$ is the ionization potential of the molecule.[3]

It should be stressed that the London dispersion forces given by Eq. (4) govern intermolecular interactions in organic liquids.[3] They are several orders of magnitudes larger than the dipole-dipole Keesom and Debye forces described by Eqs. (2)–(3).[3] The direct comparison of dipole-dipole and London interactions for the acetone molecule, featured by a dipole moment as high as 2.9 D, is supplied in the Appendix 1. This immediately explains why polar and non-polar liquids demonstrate close values of surface tension.

Indeed, $CCl_4$, which is a non-polar ($p = 0$) liquid featured by a value of surface tension similar to that of strongly polar chloroform ($p = 1.60$ D) and dichloromethane ($p = 1.04$ D), as presented in Table I. Acetone is characterized by an extremely high dipole moment of its molecule, and a low value of surface tension ($p = 2.9$ D, $\gamma = 24$ mJ/m$^2$). The data summarized in Table 1 show that there is no correlation between surface tension and dipole moment. Thus, it could be suggested that the attraction between molecules is described by Eq. (4), which is independent of the permanent dipole moment of a molecule. Taking into account: $\alpha \approx 4\pi\varepsilon_0 r_m^3$ ($r_m$ is the radius of the molecule, see Appendix 2) and substituting $\alpha$ into Eq. (4) yields:

$$U_{attr}(r) = -\frac{3}{4} I \frac{r_m^6}{r^6} = -\frac{3I}{2^8}\left(\frac{d_m}{r}\right)^6. \qquad (5)$$

The corresponding Lennard-Jones potential, considering both repulsion and attraction, is given by Eq. 6:

$$U(r) = \frac{3I}{2^9}\left[\left(\frac{d_m}{r}\right)^{12} - 2\left(\frac{d_m}{r}\right)^6\right]. \qquad (6)$$

The potential in Eq. (6) at its minimum equals:

$$|U(d_m)| = \frac{3I}{2^9}. \qquad (7)$$

Substituting Eq. (7) into Eq. (1) finally yields:

$$\gamma \cong \frac{3N}{2^{10}} \frac{I}{d_m^2}. \qquad (8)$$

Equation (8) answers the question asked in the title of the paper: Why do surface tensions of organic liquids demonstrate close values? Indeed, the surface tension of organic liquids depends on the potential of the ionization and the diameter of the molecule only. These parameters vary slightly for all organic liquids. Moreover, Eq. (8) predicts $\gamma \approx (\text{const})/d_m^2$; this dependence actually takes place for *n*-alkanes.[9]

In the spirit of the V. Weisskopf paper, natural scaling parameters could be introduced: $I = \mu R_y, d_m = \nu a_0$, where $R_y = me^4/2h^2 = 13.6 \text{ eV}$ is the Rydberg unit ($1\,\text{eV} = 1.6 \cdot 10^{-19}$ J), $a_0 = h^2/me^2 \approx 0.5$ Å (1 Å = $10^{-10}$ m) is the Bohr radius, *m* and *e* are the mass and charge of electron and *h* – is the Planck constant.[7] Substituting scaling parameters in Eq. (8) gives:

$$\gamma \cong \frac{3N}{2^{10}} \frac{R_y}{a_0^2} \frac{\mu}{\nu^2} \ . \qquad (9)$$

Actually potentials of ionization for most organic liquids are close to 10 eV; thus, $\mu \approx 0.7$ and parameter $\nu$ varies in the range of $6 < \nu < 12$.[3,5,10] Substituting $N = 6$, $\mu = 0.7, \nu = 10$ into Eq. (9) supplies for $\gamma$ the realistic value of 0.1 J/m$^2$.

It should be mentioned that the molar enthalpy of vaporization of liquids is also governed by the pair intermolecular interaction. Thus, it could be expected that it would vary slightly. Qualitative data supplied in Table I, support this suggestion. Indeed, for the discussed liquids it varies within the range of 28.6–91.7 kJ/mol.

## III. DISCUSSION

It is well accepted that there exist phenomena that have no explanation in the realm of classical physics. One such effect is the conductivity of metals. This paper shows that surface tension is also a pure quantum effect. London dispersion forces governing the surface tension originate in quantum mechanics. London dispersion forces are insensitive to the permanent dipole moment of the molecule, and depend on the potential of ionization and diameter of the molecule. These parameters vary slightly for organic liquids (with the exception of polymers and polymer solutions, where the situation is extremely complicated). Thus, the proximity of values of surface tensions of organic liquids becomes clear.

## ACKNOWLEDGEMENTS

The author is grateful to Professor Kazachkov, Professor A. Voronel, and Dr. G. Whyman for extremely fruitful discussions; to Professor M. Zinigrad for his generous support of the author's scientific and pedagogical activity, and to Mrs. Ye. Bormashenko, Mrs. O. Bormashenko and Mrs. A. Musin for their help in the preparation of the paper.

## APPENDIX 1: THE COMPARISON OF DIPOLE-DIPOLE AND LONDON ATTRACTIONS FOR THE ACETONE MOLECULE

Let us compare the dipole-dipole $U_{d-d}$ and London $U_{London}$ attractions for the acetone molecule, which is featured by a very high dipole moment $p = 2.9$ D; hence, the high value of $U_{d-d}$ is expected. Equations (2) and (4) yield:

$$\frac{U_{\text{d-d}}}{U_{\text{London}}} = \frac{4p^4}{9kT\alpha^2 I}, \qquad (10)$$

where $\alpha = 4\pi\varepsilon_0 r_m^3$, for acetone $r_m = 3.15 \cdot 10^{-10}$ m (see Ref. 11) and $I = 9.7$ eV (see Ref. 12). Remember that: 1 eV = $1.6 \cdot 10^{-19}$ J; 1 D = $3.3 \cdot 10^{-30}$ C·m; $k$ = $1.38 \cdot 10^{-23}$ J/K; $\varepsilon_0$ = $8.85 \cdot 10^{-12}$ F/m.

Substituting of this data in Eq. (10) yields for $T = 300$K:

$$\frac{U_{\text{d-d}}}{U_{\text{London}}} \cong 5 \cdot 10^{-2}.$$

We recognize that even for the acetone molecule possessing an unusually high dipole moment, the dipole-dipole interaction is much weaker than the London one.

**APPENDIX 2: POLARIZABILITY OF SPHERICAL MOLECULES**

Assume a spherical molecule consisting of a nucleus of charge $+q$ and an electron cloud of radius $r_m$ and charge $-q$. A static electric field $E$ induces a dipole moment $p$ in a molecule:

$$p = \alpha E, \qquad (11)$$

where $\alpha$ is called the polarizability of the molecule. We estimate the order of magnitude of $\alpha$ by the following considerations.[3,7] Let us estimate how strong a field is needed to displace electron cloud relative to the nucleus so much that the nucleus is moved to the rim of the cloud. The dipole moment of a molecule would be then $p = qr_m$. The Coulomb force $F$ which would drive the nucleus to the center is $F = kq^2/r_m^2$. This force is compensated by the external force $qE$, thus, we would need an electric field $E = kq/r_m^2$ to hold a molecule in this unusual state with the nucleus at the rim. Substitution in Eq. (11) yields:

$$p = qr_m = \alpha k \frac{q}{r_m^2}. \qquad (12)$$

We obtain eventually: $\alpha = r_m^3/k = 4\pi\varepsilon_0 r_m^3$ in the SI system of units.

**REFERENCES**

1. L. D. Landau and E. M. Lifshitz, *Statistical Physics* (Butterworth-Heinemann, Oxford, 2000).

2. P.G. de Gennes, Fr. Brochard-Wyart, D. Quéré, *Capillarity and Wetting Phenomena. Drops, Bubbles, Pearls, Waves* (Springer, New York, 2002).

3. H. Yildirim Erbil, *Surface Chemistry of Solid and Liquid Interfaces* (Blackwell, Oxford, 2006).

4. A. W. Adamson, *Physical Chemistry of Surfaces*, Fifth Edition, (Interscience Publishers, New York, 1990).

5. I. K. Kikoin, *Tables of Physical Values* (Atomizdat, Moscow, 1976).

6. S. Rehman, Z. F. Khan, T. Maqbool, "Physical and spectroscopic characterization of Pakistani honey," Cien. Inv. Agr. **35** (2), 199–204 (2008).

7. V. F. Weisskopf, "Of atoms, mountains, and stars: a study in qualitative physics," Science, **187** (4177), 605–612 (1975).

8. J. C. Moore, A. Kazachkov, A. G. Anders, C. Willis. "The Danger of Misrepresentations in Science Education," G. Planinsic, A. Mohoric

(eds), Informal Learning and Public Understanding of Physics', 3rd International GIREP Seminar 2005. Selected Contributions, University of Ljubljana, Ljubljana, Slovenia, 399–404 (2005).

9. Y.-Z. Su, R. W. Flumerfelt, "A continuum approach to microscopic surface tension for the *n*-alkanes," Ind. Eng. Chem. Res. **35**, 3399–3402 (1996).

10. T. M. Lowry, A. G. Nasini, "Molecular dimensions of organic compounds, Part I, General considerations," Proceedings of the Royal Society of London A, **123** (792), 686–691 (1929).

11. A. I. Toryanik, V. N. Taranenko, "Molecular mobility and structure in water-acetone solutions," J. Structural Chemistry, **28** (5), 714–719 (1988).

12. W. M. Trott, N. C. Blais, Ed. A. Walters, "Molecular beam photoionization study of acetone," J. Chem. Phys. **69**, 3150 (1978).